\documentclass{article}
\usepackage{spconf,amsmath,graphicx,hyperref,comment}

\usepackage{multirow}
\usepackage{siunitx}
\usepackage{bm}
\usepackage{amsfonts}
\usepackage{color}

\usepackage[
    backend=biber,
    style=ieee,
    maxbibnames=3,
    maxcitenames=3,
    doi=false,isbn=false,url=false
]{biblatex} 
\newcommand{\minisection}[1]{\noindent\textbf{#1}: }

\usepackage{subcaption}

\title{{Spatial-CLAP}: Learning Spatially-Aware audio--text Embeddings\\for Multi-Source Conditions}
\name{
\shortstack{
{Kentaro Seki$^{1,2}$, Yuki Okamoto$^1$, Kouei Yamaoka$^1$, Yuki Saito$^1$,}
\\
{Shinnosuke Takamichi$^{1,2}$, and Hiroshi Saruwatari$^1$}
}\thanks{\footnotesize The work was supported by 
JSPS KAKENHI Grant Number 24KJ0860, 
JST Moonshot Grant Number JPMJMS2011
JST FOREST Program JPMJFR226V,
and NII Open Collaborative Research 2025-(251S4-22735).}}
\address{$^1$ The University of Tokyo, Japan. $^2$ Keio University, Japan.\vspace{-5pt}}
\begin{document}
\ninept
\maketitle

\begin{abstract}\vspace{-4pt}
Contrastive language--audio pretraining (CLAP) has achieved remarkable success as an audio--text embedding framework,
but existing approaches are limited to monaural or single-source conditions and cannot fully capture spatial information.
The central challenge in modeling spatial information lies in multi-source conditions, where the correct correspondence between each sound source and its location is required.
To tackle this problem, we propose \textit{Spatial-CLAP}, which introduces a content-aware spatial encoder that enables spatial representations coupled with audio content.
We further propose \textit{spatial contrastive learning} (SCL), a training strategy that explicitly enforces the learning of the correct correspondence and promotes more reliable embeddings under multi-source conditions.
Experimental evaluations, including downstream tasks, demonstrate that Spatial-CLAP learns effective embeddings even under multi-source conditions, and confirm the effectiveness of SCL.
Moreover, evaluation on unseen three-source mixtures highlights the fundamental distinction between conventional single-source training and our proposed multi-source training paradigm.
These findings establish a new paradigm for spatially-aware audio--text embeddings.
\vspace{-8pt}
\end{abstract}

\begin{keywords}
Polyphonic SELD, CLAP, multiple DoA estimation, automated audio captioning
\end{keywords}

\vspace{-6pt}
\section{Introduction}
\vspace{-6pt}
\label{sec:intro}

Recent advances in audio--text embedding models such as  CLAP~\cite{wu2023large, elizalde2023clap}, AudioCLIP~\cite{guzhov2022audioclip}, Pengi~\cite{deshmukh2023pengi}, and SALMON~\cite{tang2023salmonn} have enabled learning a shared representation space for audio and text.
These models reliably capture \textit{content information}---{what sound events are occurring}---and transfer well to various downstream tasks, including retrieval~\cite{wu2023large}, captioning~\cite{ghosh2024recap}, and target speech extraction~\cite{seki2025language}. 
However, most existing approaches are designed for monaural audio, which inevitably discards most of \textit{spatial information}---{where acoustic events are occurring}.
Spatial information plays a crucial role in human binaural hearing, enabling listeners to localize acoustic events~\cite{carlini2024auditory};
thus, our auditory perception leverages both content information and spatial information to understand complex acoustic scenes.
This fact underscores the importance of realizing audio--text embeddings that integrate both types of information.

\begin{figure}[t]
  \centering
  \begin{subfigure}[t]{1.0\linewidth}
    \centering
    \includegraphics[width=\linewidth]{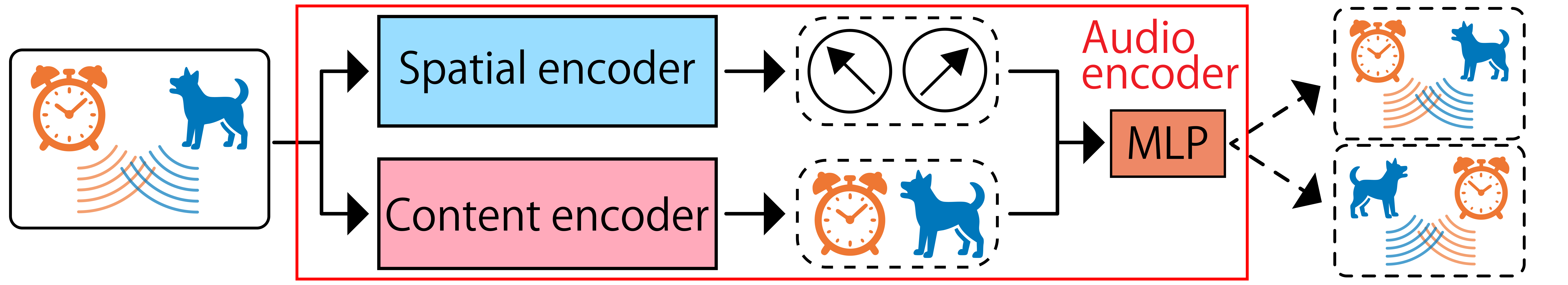}
    \vspace{-14pt}
    \caption{Conventional}
      \label{fig1:a}
    \vspace{3pt} 
  \end{subfigure}

  \begin{subfigure}[t]{1.0\linewidth}
    \centering
    \includegraphics[width=\linewidth]{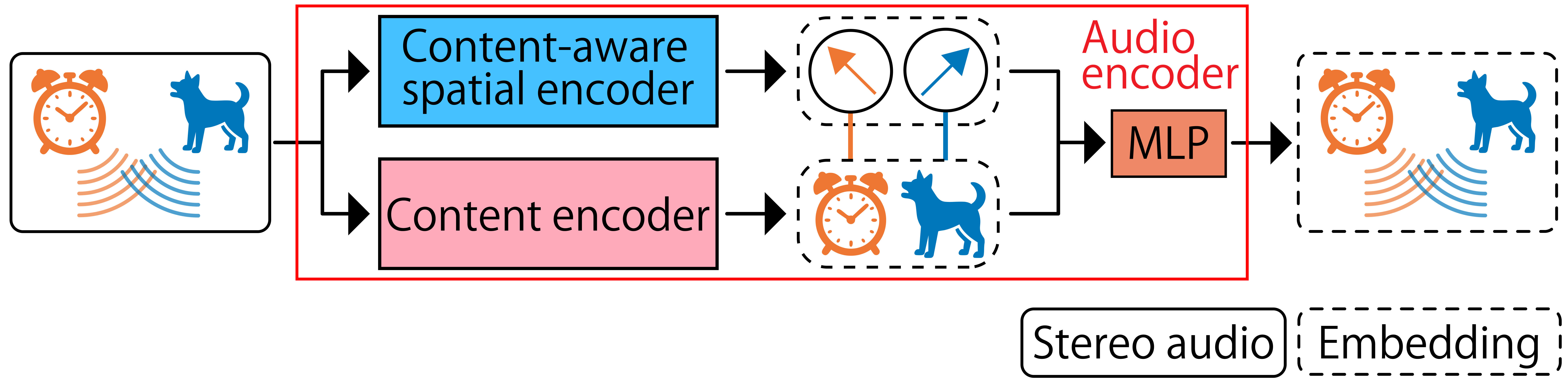}
    \vspace{-26pt}
    \caption{Spatial-CLAP (Ours)
    \;\;\;\;\;\;\;\;\;\;\;\;\;\;
    }
      \label{fig1:b}
  \end{subfigure}
  \vspace{-15pt}
\caption{Comparison of audio encoders. (a) Conventional method~\cite{devnani2024learning} encodes content and spatial information separately, causing permutation problem under multi-source conditions. (b) Spatial-CLAP introduces a content-aware spatial encoder aligning content and spatial embeddings.}
  \label{fig1}
\vspace{-13pt}
\end{figure}

A prior spatial extension of CLAP introduced both a content encoder and a spatial encoder to incorporate spatial information into audio--text embeddings~\cite{devnani2024learning}.
While this approach successfully integrates spatial information in single-source conditions, it fails to generalize to multi-source conditions.
As illustrated in Fig.~\ref{fig1:a}, when content information and spatial information remain unaligned, it leads to a permutation problem in establishing the correct \textit{content--space correspondence} across multiple sources.
As a result, the embeddings fail to capture which event occurs where, and thus cannot be regarded as a valid representation of the acoustic scene.
This problem raises the broader challenge of embedding design:
\textit{Does a unified representation of content and spatial information truly exist under multi-source conditions?}

To address this challenge, we propose \textit{Spatial-CLAP}, a stereo audio--text embedding model that unifies content and spatial information within a single fixed-dimensional embedding, even under multi-source conditions.
As shown in Fig.~\ref{fig1:b}, Spatial-CLAP introduces a \textit{content-aware spatial encoder (CA-SE)} that learns spatial embeddings coupled with content information, enabling robust embedding learning under multi-source conditions.
Unlike prior work, Spatial-CLAP is explicitly trained under multi-source conditions.
Furthermore, we propose \textit{spatial contrastive learning (SCL)}, which leverages negative examples with swapped spatial attributes to enforce content--space correspondence.
Our extensive experiments include evaluation on a downstream task named \textit{spatial audio captioning}, which aims to generate an audio caption describing both content information and spatial information.
Experimental results demonstrate that Spatial-CLAP effectively captures both content and spatial information, with SCL improving content--space correspondence and promoting reliable representations under multi-source conditions. 
Also, the evaluation with unseen three-source mixtures further highlights the fundamental difference between conventional single-source training and our multi-source training paradigm.
To ensure full reproducibility and to establish a baseline for future research on spatial audio--text embeddings, we release our code and pretrained models.\footnote{\label{fn:github}\url{https://github.com/sarulab-speech/SpatialCLAP}}

\section{Related Work}
\label{sec:related_work}
\vspace{-3pt}

Our study relates to polyphonic sound event localization and detection (SELD)~\cite{adavanne2018sound}, which estimates sound event activity together with the direction of arrival (DoA) for each source in multi-source conditions.
This task explicitly handles the content--space correspondence, associating detected event activity (content information) with its corresponding DoA (spatial information).
Representative approaches include multitask learning~\cite{adavanne2018sound}, unified vector representations named ACCDOA~\cite{shimada2021accdoa}, and track-wise formulations with permutation-invariant training~\cite{cao2021improved, shimada2022multi}.
However, these methods inherently depend on predefined class labels, which limits their flexibility in open-vocabulary settings.

To go beyond predefined class labels, embed-ACCDOA~\cite{shimada2024zero} leveraged CLAP-based embeddings for open-vocabulary SELD.  
This model adopts a track-wise format, that is, it outputs an embedding for each source, given a fixed number of output tracks.
However, the text modality is not restricted to such a predefined-track structure; rather, its strength lies in flexibility for describing arbitrary content.
Since our goal is to align audio with this flexible modality, we adopt a fixed-dimensional embedding that is not restricted to any predefined structure, thereby better matching the nature of language.
The conventional spatial extension of CLAP~\cite{devnani2024learning} also shares a similar motivation with ours, but as discussed in Section~\ref{sec:intro}, it does not tackle the fundamental difficulty of handling spatial information in multi-source conditions.

\vspace{-8pt}
\section{Method}
\label{sec:method}
\vspace{-4pt}
Spatial-CLAP consists of two encoders: 
an \textit{audio encoder} that takes a stereo audio signal as input, 
and a \textit{text encoder} that takes a caption as input. 
Both encoders output fixed-dimensional embeddings, 
which are aligned within a shared space through a contrastive learning framework.

\vspace{-8pt}
\subsection{Model Architecture}
\vspace{-5pt}

As shown in Fig.~\ref{fig1:b}, our audio encoder processes a stereo signal by integrating two encoders.
The \textit{content encoder} (CE) produces one fixed-dimensional embedding by feeding the average of the left and right channels into the monaural CLAP encoder.
This design allows Spatial-CLAP to leverage the existing monaural CLAP pretrained on large-scale audio--text data, enabling rich and fine-grained representations of content information.
The \textit{content-aware spatial encoder} (CA-SE) is adapted from a model designed for SELD, and it takes a stereo signal to produce a fixed-dimensional embedding.
Pretraining on the SELD task enables CA-SE to learn spatial embeddings coupled with content information, thereby facilitating robust modeling of content--space correspondence in multi-source conditions.
The outputs of the CE and CA-SE are concatenated and processed through a two-layer multilayer perceptron (MLP) to form the final audio embedding.
By combining the CE and CA-SE, Spatial-CLAP achieves representations that unify rich content information with accurate content--space correspondence, thus providing effective representations under multi-source conditions.

For the text encoder, we adopt a model pretrained on large and diverse text corpora to produce fixed-dimensional embeddings, and fine-tune it in our contrastive learning framework, following the monaural CLAP setting~\cite{wu2023large, elizalde2023clap}.

\newenvironment{Equation}
  {\vspace{-5pt}\begin{equation}}
  {\end{equation}\vspace{-15pt}\\\noindent}
  
\vspace{-6pt}
\subsection{Contrastive Learning Framework}
\vspace{-4pt}
As in the monaural CLAP setting~\cite{wu2023large, elizalde2023clap}, we employ a contrastive learning framework with the standard in-batch InfoNCE loss~\cite{radford2021learning}, which aligns audio and text embeddings into a shared space.
We train Spatial-CLAP on stereo audio-caption pairs that include multi-source conditions and spatial descriptions in the captions, allowing the embeddings to capture both content and spatial information.
However, content--space correspondence in multi-source conditions is inherently complex, and simple contrastive learning may be insufficient to fully capture it.
  
\begin{figure}[t]
    \centering
    \includegraphics[width=0.95\linewidth]{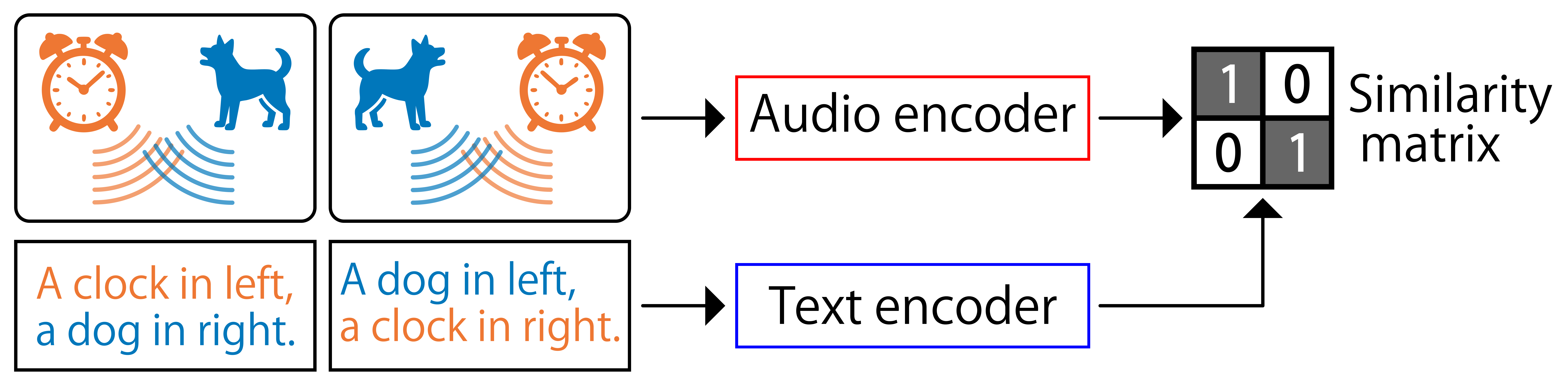}
    \vspace{-8pt}
    \caption{
Spatial contrastive learning (SCL) enforces the model to explicitly learn content--space correspondence in multi-source environments 
by adding permuted content--space assignment audio--text pairs as hard negative examples.
}
    \label{fig:scl}
    \vspace{-15pt}
\end{figure}

To address this issue, we introduce \textit{spatial contrastive learning} (SCL). 
As illustrated in Fig.~\ref{fig:scl}, SCL introduces hard negative examples by permuting content--space correspondence to explicitly enforce the model to learn the correct correspondence.
To formalize this idea, a two-source stereo signal is denoted as
\begin{Equation}
    \bm y(t) = (\bm h_1 * x_1)(t) + (\bm h_2 * x_2)(t),
\end{Equation}
where $x_1(t)$ and $x_2(t)$ are the sources, $\bm h_1(t)$ and $\bm h_2(t)$ are the corresponding room impulse responses (RIRs), and $*$ denotes convolution. 
SCL further constructs a sample with permuted content--space correspondence:\footnote{%
In general, SCL generates $n!-1$ examples for $n$ sources.}
\begin{Equation}
    \bm y_{\text{SCL}}(t) = (\bm h_2 * x_1)(t) + (\bm h_1 * x_2)(t).
\end{Equation}
Both $\bm y(t)$ and $\bm y_{\text{SCL}}(t)$ with their corresponding captions are included in the same batch, where the permuted pairs act as hard negative examples. 
In this way, SCL provides explicit supervision for content--space correspondence, making the learned embeddings robust under multi-source conditions.

\vspace{-6pt}
\section{Experimental evaluation}
\label{sec:experiment}
\vspace{-4pt}

\begin{table*}
\centering
\footnotesize
\caption{
Results of embedding-based evaluation.
``Ours'' consistently shows strong performance across all metrics.
}
 \vspace{-8pt}
 \label{tab:result}
 \begin{tabular}{|c||cc|cc|cc|cc|c|}
  \hline  
  \multirow{2}{*}{Method} & 
  \multicolumn{2}{c|}{R@1 (fixed-RIR)} &
  \multicolumn{2}{c|}{R@1 (1-src)} &
  \multicolumn{2}{c|}{R@1 (2-src)} &
  \multicolumn{2}{c|}{Spatial classification} &
  Content-space
  \\
   & A2T & T2A & A2T & T2A  & A2T & T2A 
   & 1-src & 2-src & assignment
  \\ \hline \hline
  (Random) &
  $(0.10\%)$ & $(0.10\%)$ & $(0.02\%)$ & $(0.02\%)$ & $(0.01\%)$ & $(0.01\%)$ &
  $(20\%)$ & $(4\%)$ & $(50\%)$  \\ \hline
Monaural & $22.02\%$ & $\mathbf{21.60\%}$ & $4.81\%$ & $5.33\%$ & $1.66\%$ & $2.07\%$ & $22.33\%$ & $7.00\%$ & $51.85\%$ \\
\hline
Conventional & $16.87\%$ & $16.67\%$ & $16.60\%$ & $17.10\%$ & $4.16\%$ & $4.98\%$ & $96.40\%$ & $22.43\%$ & $48.77\%$ \\
Structured & $17.28\%$ & $17.90\%$ & $17.12\%$ & $\mathbf{18.89\%}$ & $4.22\%$ & $4.85\%$ & $\mathbf{99.28\%}$ & $23.66\%$ & $52.26\%$ \\
\hline
Ours & $\mathbf{23.25\%}$ & $19.24\%$ & $\mathbf{20.78\%}$ & $18.46\%$ & $\mathbf{20.79\%}$ & $\mathbf{21.34\%}$ & $98.56\%$ & $\mathbf{61.32\%}$ & $\mathbf{81.69\%}$ \\
Ours (w/o SCL) & $20.47\%$ & $19.44\%$ & $20.33\%$ & $18.85\%$ & $18.26\%$ & $18.43\%$ & $98.46\%$ & $57.82\%$ & $74.49\%$ \\
Ours (w/o CLAP) & $6.48\%$ & $5.25\%$ & $6.36\%$ & $5.84\%$ & $6.06\%$ & $6.16\%$ & $95.88\%$ & $60.70\%$ & $77.37\%$ \\
Ours (w/o SELD) & $17.59\%$ & $17.80\%$ & $18.19\%$ & $17.08\%$ & $15.04\%$ & $16.42\%$ & $95.47\%$ & $52.26\%$ & $71.60\%$
 \\ \hline
 \end{tabular}
 \vspace{-12pt}
\end{table*}

\subsection{Experimental Setup}
\vspace{-3pt}
\subsubsection{Datasets}
\label{sec:data}
\vspace{-3pt}
We constructed stereo audio signals by convolving the monaural audio with simulated RIRs, and augmented the captions with spatial descriptions. 
For multi-source scenes, multiple stereo signals were summed to form mixtures, and their captions were concatenated. 

The monaural audio paired with captions is from AudioCaps 2.0~\cite{audiocaps}, 
consisting of $91{,}256$ training, $2{,}475$ validation, and $975$ test samples. 
Although each audio clip may contain mixtures of different sounds, we treated each clip as a single source in the spatial sense, since the signal is monaural without spatial separation.
The RIRs were simulated using pyroomacoustics~\cite{scheibler2018pyroomacoustics}, 
and the spatial descriptions indicate DoA, chosen from five classes 
(``front-left,’’ ``front,’’ ``front-right,’’ ``left,’’ ``right’’). 
Since the two-microphone setting cannot distinguish front from back, the DoA labels were folded. 
We simulated $440$ reverberant rooms with reverberation times ranging from $130~\si{ms}$ to $260~\si{ms}$ at a sampling rate of $16$~\si{kHz}, split into $360$ for training, $60$ for validation, and $60$ for testing, and ensured that these sets used disjoint room configurations.
Further implementation details are provided in our public repository.\footref{fn:github}

\vspace{-3pt}
\subsubsection{Model Configuration}
The CE adopted HTS-AT~\cite{chen2022hts}, 
initialized with CLAP pretraining~\cite{laion_clap}, and produced a $768$-dimensional embedding. 
For the CA-SE, we pretrained SELDNet~\cite{adavanne2018sound} on the data described in Section~\ref{sec:data}. 
SELDNet takes magnitude and phase spectrograms of multi-channel short-time Fourier transforms as input, applies convolutional and recurrent layers, and predicts sound event activity and DoA at the frame level through task-specific MLPs.
The model is pretrained with frame-wise cross-entropy for activity and mean squared error for DoA.
When adapting SELDNet as CA-SE, we replace the output MLPs with temporal average pooling, yielding a 512-dimensional embedding.
These CE and CA-SE embeddings were concatenated and passed through a two-layer MLP 
(with $512$ hidden units and ReLU activations) to produce the final $512$-dimensional audio embedding. 
For the text encoder, we adopted RoBERTa-base~\cite{liu2019roberta}.

\vspace{-4pt}
\subsubsection{Training Details}
\vspace{-2pt}
We trained all model parameters of Spatial-CLAP (including the pretrained CE, CA-SE, and RoBERTa) using Adam~\cite{kingma2014adam} 
with a learning rate of $1\times10^{-5}$, a batch size of $128$, and for $50$ epochs. 
Each batch consisted of $64$ single-source samples and $64$ samples obtained by applying SCL to $32$ two-source pairs, 
and the InfoNCE loss~\cite{radford2021learning} was computed over all $128$ samples.

\vspace{-4pt}
\subsubsection{Compared Methods}
\vspace{-2pt}
We compared the following model configurations and training strategies of audio–text embedding models (see Fig.~\ref{fig1}):
\begin{itemize} \vspace{-1.0mm} \itemsep 0mm \leftskip -5mm
    \item \textbf{Monaural}: To correspond to the conventional CLAP setting, we consider a Spatial-CLAP without CA-SE. 
    \item \textbf{Conventional}: As a baseline faithfully reproducing the prior work~\cite{devnani2024learning}, we pretrained a spatial encoder (SE) on DoA estimation, and trained Spatial-CLAP under single-source conditions.
    The SE shares the same architecture as our CA-SE, but since it is not coupled with content information, we refer to it simply as SE rather than CA-SE. 
    \item \textbf{Structured}: As a design that cannot learn content--space correspondence, 
        we replaced CA-SE with the SE used in ``Conventional,''
        and replaced the final MLP of the audio encoder with independent MLPs applied separately to SE and CE.
        SE and CE respectively produced $64$- and $448$-dimensional embeddings, 
        which were concatenated to form the final $512$-dimensional audio embedding. 
    \item \textbf{Ours}: Our proposed Spatial-CLAP.
    ``Ours (w/o XX)'' denotes ablation variants:
    ``Ours (w/o SCL)'' excludes SCL,
    ``Ours (w/o CLAP)'' does not initialize the CE with CLAP,
    and ``Ours (w/o SELD)'' replaces the CA-SE with the SE from ``Conventional.''
\end{itemize}

\vspace{-4pt}
\subsection{Evaluation}

\subsubsection{Embedding-Based Evaluation}
\label{sec:eval:emb}

\minisection{R@1 score} 
To test whether Spatial-CLAP learns aligned representations of audio and text, we measured recall@1 (R@1) score~\cite{wu2023large} for audio-to-text (A2T) and text-to-audio (T2A) retrieval, measured as the percentage of queries with the correct pair ranked at the top. 
%
The \textit{fixed-RIR} setting adopted a single pre-selected RIR,
focusing on content information exclusively.
The \textit{1-src} setting considers a single-source condition with all five DoA classes, 
focusing on both content and spatial information.  
The \textit{2-src} setting considers a two-source condition with all $5 \times 5 = 25$ combinations of DoA classes,
thereby assessing generalization ability 
to multi-source conditions.  

\minisection{Spatial classification} 
To directly evaluate the ability of Spatial-CLAP to represent spatial information,
we conducted a spatial classification task by selecting the caption with the highest embedding similarity to the audio among its candidates. 
In the \textit{1-src} setting, the candidate set consists of five spatial descriptions (e.g., ``front-left''), 
whereas in the \textit{2-src} setting, it consists of the same five descriptions, together with 10 distinct pairs of two descriptions (e.g., ``front-left'' + ``front-right'').

\minisection{Content--space assignment} 
For two-source mixtures, we evaluated whether the model can correctly learn content--space correspondence.
For each mixture, we calculated the embedding similarity between the audio and the correct caption, 
and that between the audio and a caption with incorrect content--space correspondence.
We then measured the proportion of cases where the former was higher. 

\vspace{-4pt}
\subsubsection{Spatial audio captioning}
\vspace{-2pt}
We also evaluated the embeddings on a downstream task named \textit{spatial audio captioning}, where the goal is to generate captions including spatial descriptions.  

\minisection{Model and training}
We froze the Spatial-CLAP audio encoder and linearly projected its embeddings into $10$ tokens, 
which were used as prefix inputs to a GPT-2~\cite{radford2019language} decoder. 
The projection layer and GPT-2 were trained on the data described in Section~\ref{sec:data} for $50$ epochs using AdamW~\cite{loshchilov2017decoupled} (learning rate: $1\times 10^{-5}$, batch size: $128$).

\minisection{Evaluation} 
We adopted evaluation metrics commonly used in audio captioning~\cite{mei2022automated}, including BLEU~\cite{papineni2002bleu}, ROUGE-L~\cite{lin2004rouge}, METEOR~\cite{banerjee2005meteor}, CIDEr~\cite{vedantam2015cider}, SPICE~\cite{anderson2016spice}, SPIDEr~\cite{liu2017improved}, BERTScore~\cite{zhang2019bertscore}, and SentenceBERT (SBERT)~\cite{reimers2019sentence}.

In addition, we introduced two spatially-oriented metrics.  
\textit{Direction-wise SBERT (DW-SBERT)} decomposes a caption according to spatial classes and computes SBERT similarity for each spatial class, averaging the results. 
Since the score decreases when the directional description attached to each source is incorrect, DW-SBERT works as a metric that jointly evaluates both source content and spatial location.
\textit{Spatial description} measures how often the generated captions contain the correct spatial description. 

{
\setlength{\tabcolsep}{5pt} 
\begin{table*}
\centering
\footnotesize
\caption{Evaluation results on spatial audio captioning, 
which aims to generate captions containing spatial descriptions. 
}
\vspace{-2pt}
 \label{tab:captioning}
 \begin{tabular}{|c||cccccccc|cc|}
\hline  
Method & BLEU & ROUGE-L & METEOR & CIDEr & SPICE & SPIDEr &
BERTScore & SBERT & DW-SBERT & Spatial desc.
\\ \hline \hline
Monaural & $0.0735$ & $0.2823$ & $0.1789$ & $0.1986$ & $0.1757$ & $0.1871$ & $0.3769$ & $0.5520$ & $0.2196$ & $0.1770$
\\ \hline
Conventional & $0.1329$ & $0.3497$ & $0.1984$ & $0.2075$ & $0.2416$ & $0.2246$ & $0.3898$ & $0.5026$ & $0.3620$ & $0.6955$
\\ 
Structured & $0.1323$ & $0.3487$ & $0.1997$ & $0.2154$ & $0.2418$ & $0.2286$ & $0.3899$ & $0.5137$ & $0.3630$ & $0.6461$
\\ \hline
Ours & $\mathbf{0.1463}$ & $\mathbf{0.3709}$ & $\mathbf{0.2135}$ & $\mathbf{0.2553}$ & $\mathbf{0.2658}$ & $\mathbf{0.2606}$ & $\mathbf{0.4152}$ & $\mathbf{0.5564}$ & $\mathbf{0.4144}$ & $\mathbf{0.7942}$
\\
Ours (w/o SCL) & $0.1455$ & $0.3685$ & $0.2121$ & $0.2482$ & $0.2589$ & $0.2536$ & $0.4118$ & $0.5456$ & $0.4071$ & $0.7922$
 \\ \hline
 \end{tabular}
 \vspace{-7pt}
\end{table*}

}

\vspace{-4pt}
\subsection{Results}
\label{sec:results}

\subsubsection{Embedding-Based Evaluation}
\label{sec:eval:emb}
Table~\ref{tab:result} shows the results described in Section~\ref{sec:eval:emb}.  
``Monaural'' cannot handle spatial information, and thus its spatial classification accuracy remained at the chance level.  
``Conventional'' can capture spatial information and thereby improves the spatial classification accuracy (1-src), but its R@1 score (2-src), spatial classification accuracy (2-src) and content--space assignment accuracy remain low, indicating that it cannot handle multi-source conditions.
Furthermore, ``Structured'' outperformed ``Conventional'' on most metrics, highlighting the limitations of ``Conventional.''

In contrast, ``Ours'' achieved the best results across all evaluations under the 2-src setting (R@1 score (2-src), spatial classification accuracy (2-src), and content--space assignment), indicating Spatial-CLAP successfully learns embeddings well adapted to multi-source conditions.
Furthermore, ``Ours'' also demonstrated competitive results in content-only evaluations (R@1 score (fixed-RIR)) and in the 1-src setting, highlighting its broad applicability across different scenarios.

When compared to ``Ours (w/o SCL),'' ``Ours'' shows consistent improvements across all 2-src evaluation metrics, demonstrating that SCL enables Spatial-CLAP to learn more reliable embeddings under multi-source conditions.
``Ours (w/o CLAP)'' suffered severe drops in R@1 scores, confirming the contribution of the pretrained monaural CLAP model in leveraging rich content representations.
Finally, ``Ours (w/o SELD)'' exhibited degraded performance in both the 1-src and 2-src settings, which validates the role of the CA-SE in effectively coupling spatial information with content information.

\subsubsection{Spatial Audio Captioning}
Table~\ref{tab:captioning} presents the results of spatial audio captioning.
Among the ablation models, we focused on ``Ours (w/o SCL)'' due to its strong R@1 performance.

As a result, ``Ours'' achieved the highest scores across all metrics, demonstrating that its audio embeddings can be effective for downstream tasks.
However, the observed performance gaps between methods varied depending on the evaluation metric.
In particular, under the SBERT evaluation, ``Monaural'' and ``Ours'' achieved almost identical scores, 
although ``Monaural,'' which cannot handle spatial information, inherently lacks this information.
This is because SBERT primarily measures semantic similarity and therefore tends to place more weight on content information while largely overlooking spatial information.
In contrast, under DW-SBERT, ``Monaural,'' which cannot represent spatial information, obtained significantly lower scores than ``Ours.''
These findings suggest that conventional semantic-based metrics are insufficient for adequately assessing spatial aspects of captions, 
highlighting the necessity of spatially-aware metrics such as DW-SBERT to properly evaluate spatial audio captioning. 

\subsubsection{Visualization of embeddings}

\begin{figure}[t]
  \centering
  \begin{minipage}[t]{0.27\linewidth} 
    \centering
    \includegraphics[width=1.0\linewidth]{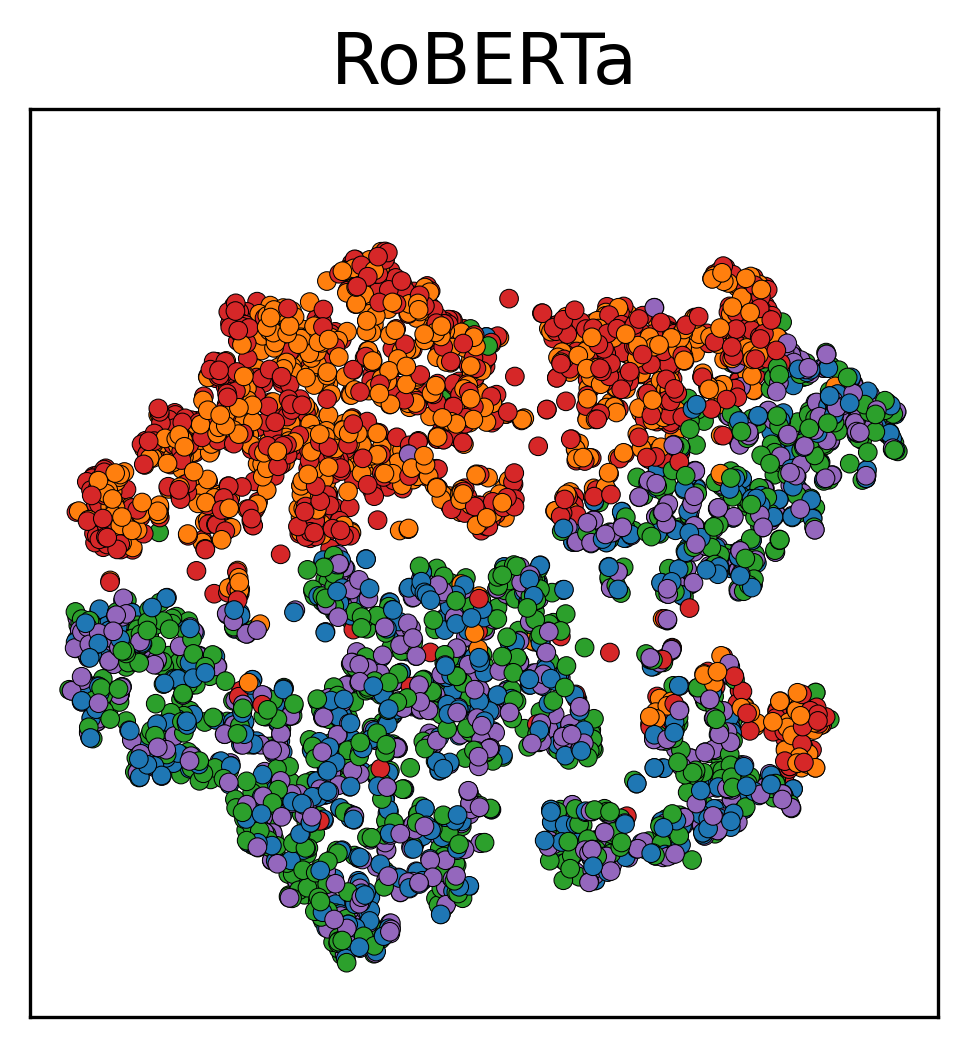}
    \vspace{-15pt}
    \subcaption{RoBERTa~\cite{liu2019roberta}}
      \label{fig:embedding:roberta}
  \end{minipage}
  \hfill
  \begin{minipage}[t]{0.72\linewidth} 
    \centering
    \includegraphics[width=1.0\linewidth]{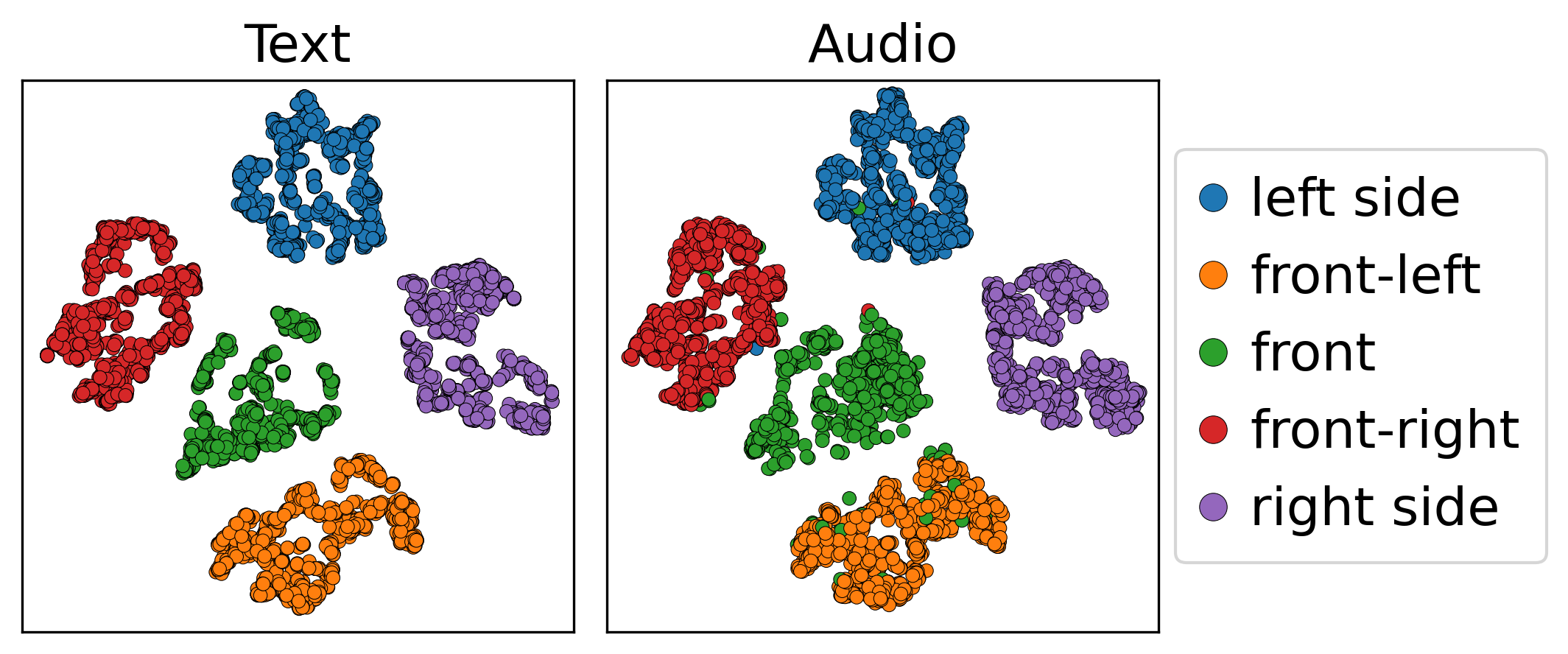}
    \vspace{-15pt}
    \subcaption{``Ours''
    \;\;\;\;\;\;\;\;\;\;\;\;\;\;\;}
    \label{fig:embedding:ours}
  \end{minipage}
  \vspace{-8pt}
  \caption{Comparison of t-SNE~\cite{maaten2008visualizing} visualizations of embeddings.
While RoBERTa produces mixed clusters without clear separation,
``Ours'' forms distinct clusters.}
  \vspace{-5pt}
  \label{fig:embedding}
\end{figure}

Fig.~\ref{fig:embedding} shows the t-SNE~\cite{maaten2008visualizing} visualization of audio and caption embeddings under the single-source condition for both RoBERTa~\cite{liu2019roberta} and ``Ours.''
The embeddings were obtained from the same dataset used for the R@1 evaluation (1-src).
While RoBERTa embeddings do not exhibit clear separation across spatial classes, 
``Ours'' produces distinct clusters in both text and audio embeddings.
These results suggest that these structures emerged through audio--text contrastive learning with alignment to audio. 
This may be due to spatial information being only weakly represented in the text embeddings, while alignment with audio emphasized it and revealed a clear cluster structure. 

\subsection{Generalization to three-source conditions}
We evaluated the content--space assignment under the three-source condition.  
Among the six possible content--space assignments, only the one with all correspondences correct was regarded as the ground truth, 
and we measured the proportion of times the model successfully selected it.
As a result, ``Conventional'' remained at $16.31\%$, which is close to the random chance rate of $16.67\%$.
In contrast, ``Ours (w/o SCL)'' and ``Ours'' achieved $34.31\%$ and $41.77\%$, respectively, although their training data did not include the three-source condition. 
These results suggest that training with two sources enables generalization to scenarios with more sources.
These findings highlight the fundamental distinction between learning spatially-aware audio--text embeddings from single-source and from multi-source signals.

\section{Conclusion}
We introduced \textit{Spatial-CLAP}, a stereo audio--text embedding model designed to be effective even under multi-source conditions. 
The key challenge lies in establishing content--space correspondence, which we addressed by introducing a content-aware spatial encoder. 
Furthermore, we proposed \textit{spatial contrastive learning} (SCL) to encourage the model to learn the correct content--space correspondence. 
Through experiments, including evaluation on downstream tasks, we demonstrated that Spatial-CLAP learns embeddings effectively even under multi-source conditions, with the effectiveness of SCL.
These results provide a foundation for future research on spatially-aware audio--text embedding models. 
Future work includes extending our framework to more complex scenarios, such as moving sound sources.

\newpage
\printbibliography

\end{document}